\documentclass[%
 reprint,
 amsmath,amssymb,
 aps,
]{revtex4-2}

\usepackage{graphicx}% Include figure files
\usepackage{dcolumn}% Align table columns on decimal point
\usepackage{bm}
\usepackage{blindtext}
\usepackage[a4paper,
            bindingoffset=0.2in,
            left=0.75in,
            right=0.75in,
            top=1in,
            bottom=1in,
            footskip=.25in]{geometry}
\usepackage[english]{babel}		
\usepackage[utf8]{inputenc}										% Escrita de caracteres acentuados e cedilhas - 1
\usepackage[T1]{fontenc}										% Escrita de caracteres acentuados + outros detalhes técnicos fundamentais
\usepackage{amsmath,amsfonts,amssymb,amsthm,cancel,siunitx,
calculator,calc,mathtools,empheq,latexsym}
\usepackage{tikz,float}		            % Packages de figuras. 
\usepackage{booktabs,multirow,tabularx,array}          % Packages para tabela
\usepackage{natbib}
\usepackage{tikz}
\usepackage{pgfplots}
\pgfplotsset{compat=1.15}
\usepackage{mathrsfs}
\usetikzlibrary{arrows}
\usepackage{comment}
\usepackage{physics}

\newtheorem{theorem}{Theorem}

\usepackage{mathtools}

\usepackage{hyperref}
\hypersetup{colorlinks=true, allcolors=blue}
\usepackage{tikz}
\usetikzlibrary{decorations.pathmorphing,patterns}
\usepackage{pgfplots}
\pgfplotsset{compat=1.15}
\usepackage{mathrsfs}
\usetikzlibrary{arrows, arrows.meta}
\usetikzlibrary{calc,patterns,angles,quotes}
\usepackage{comment}
\usepackage{physics}
\usepgflibrary{patterns.meta} % LATEX and plain TEX and pure pgf
\usepgflibrary[patterns.meta] % ConTEXt and pure pgf
\usetikzlibrary{patterns.meta} % LATEX and plain TEX when using TikZ
\usetikzlibrary[patterns.meta] % ConTEXt when using TikZ

\begin{document}

\preprint{APS/123-QED}

\title{Creation and annihilation operators in Schrödinger representation in curved backgrounds}% Force line breaks with \\

\author{López Daniel}
\email{da.lopez12@uniandes.edu}
\affiliation{%
 Universidad de los Andes
}

\date{\today}% It is always \today, today,

\begin{abstract}
I propose modified set of creation and annihilation operators for the Schrödinger representation which is compatible with the Fock representation which differs from previous works. I take into account the relation between different non-unitary vacuums obtained from restricted frameworks like the relation between the Minkowski and Unruh vacuums. 
\end{abstract}

\maketitle

%\tableofcontents

\section{\label{sec:level1}Introduction}

Background independent Quantum Field Theories that are heavily based on the use of the Schrödinger representation has been developed in the last 30 years\cite{Kanatchikov:2019mfd, COLOSI201765, PhysRevD.70.064037, Long:1996bk, Long:1996wf}.

The understanding of this representation in curved space-time and its relation with other representations through the Algebraic Quantum Field Theory (AQFT) using the Gel’fand, Naimark, Segal (GNS) construction theorem has been studied in papers such as \cite{affine, onTheSc, Corichi2002OnTR}.  The AQFT framework has been proven more fundamental and general, making it a powerful tool to study other representations and the relations among them. The Schrödinger representation seems to be a suitable tool for finding a background independent formulation of Quantum Field Theory which can be a relevant tool in the pursuit of Quantum Gravity \cite{PhysRevD.70.064037}. In an attempt to extend the formalism in curved backgrounds \cite{Long:1996wf} (and \cite{Long:1996bk}) built the Schrödinger representation on a curved manifold in order to avoid the usual description of the Fock representation, which interprets the vacuum as a non-particle state. They argue that such an approach is inherently problematic in curved space-time, where there is no uniquely favored mode of decomposition and no guarantee that the usual concept of a particle is a good description of the spectrum of the theory. We agree with this interpretation of the vacuum. Unlike Long and Long's (1996) interpretation, Schrödinger's representation provides an intrinsic description of the vacuum. For Long and Long, the vacuum depends explicitly on quantities that are linked to the choice of the foliation. A suitable choice of modes is made by choosing a time-like Killing vector which induces a decomposition mode. 

Canonical quantization is introduced in \cite{Wald:1995yp}, where he prefers to construct the theory by using the Dirac quantization method. Here, the selection of modes are represented by the selection of the complex structure $J$. The complex structure $J$ is part of a projection operator $K=(1-iJ)/2$ which maps the phase space to a complex one which is dense in the Hilbert space. Additionally, $J$ induces a norm on a Hilbert space. Specifically, the elements $K\phi$ are identified as the positive frequency modes. Finally, he constructs a Fock representation using such a structure.

The development of a more formal and mathematical precise treatment in AQFT results in a natural relationship between different representations. This is made though the use of representation independent algebraic states in AQFT, which leads to a connection between the Fock and Schrödinger representations. This relationship was discussed in the paper \cite{Corichi2002OnTR} using the GNS theorem. They found that the momentum density operator in its functional representation should be modified as $\hat{\pi}[g] = \int_\Sigma d^3x\left(-ig\frac{\delta}{\delta\phi}+\phi (iB^{-1}-B^{-1}A)g\right)$ so both representations can match.

\cite{affine} reproduces such results by using a geometric quantization approach.  The formalism used there allows for a certain degree of freedom in the choice of the final shape. This compatibility requires a Gaussian integration measure; therefore, the states $\Psi[\phi]$ will not longer be integrated with a "Lebesgue like" measure $\mathcal{D}\phi$. Instead, it will be a "well" defined Gaussian measure $d\mu_{B^{-1}}$ which depends on operator $B$ which is one of the components of the complex structure. Therefore, the integration and normalization of the states depends strongly on the choice of $J$. The creation and annihilation operators are also modified since they depend on $\pi$. For instance, the annihilation operator is $\hat{b}_{\text{Gauss}}[\overline{\textbf{K}\lambda}]=\frac{i}{2}\int_\Sigma d^3x\:(Ag+Bh+ig)\frac{\delta}{\delta\phi}$, and the associated vacuum state is $\Psi=e^{i\theta}$, with $\theta$ being an arbitrary phase. Such a fact is not consistent with relations found among different vacuums found in Thermofield Dynamics for the Minkowski and Rindler's vacuums. Where although the relation being non-unitary \footnote{The vacuum in Minkowski cannot be written in terms of elements in the Hilbert space induced by the Rindler's modes. The coefficients goes to zero} exists. Furthermore, the explicit dependency of the vacuum in the selection of modes is lost. In this article, we modify the calculations made in \cite{Corichi2002OnTR} and \cite{onTheSc} slightly in order to obtain a desirable Gaussian vacuum which depends explicitly on the choice of $J$. Schrödinger's representation is compatible with Fock's representation in the Weyl sense. Parallel to this result, an alternative representation of a momentum density operator is obtained with its corresponding creation annihilation operators. Such representation seems to fulfill the requirements of Thermofield dynamics, especially when we attempt to relate the Minkowski vacuum to Rindler's vacuum.

\section{Complex structure J.}\label{sec:2}
We will work with an arbitrary foliation of a manifold with topology $\mathcal{M}=\mathbb{R}\times \Sigma$. The phase space $\mathbb{M}$ is composed by the cotangent bundle whose elements are pairs of the form $(\phi, \pi)$ where each element compactly supported. An element in the solution space $\mathbb{S}$ has a one to one relationship with $\phi$ in the phase space. 
The operators $K$ and $\overline{K}$ act as a projectors from the phase space $\mathbb{M}$ to the complex spaces $\mathbb{W}$ and $\overline{\mathbb{W}}$, respectively. The phase space related to complex solutions of the Klein-Gordon equation is then the disjoint sum $V_\mathbb{C}=\mathbb{W}\oplus\overline{\mathbb{W}}$. The projector operator can be written in terms of an operator $J$ through the relation $K=(1-iJ)/2$. The condition of projector $K^2=K$ implies $J^2=-1$; thus, $J$ is a complex structure. The complex structure $J$ can be written as a matrix of operators of dimension two, namely:
\begin{equation}
    -J=\mqty(
        A & B \\
        C & D
    )
\end{equation}
By the restriction of $J^2=-1$,  a set of relations among the components can be established.
\begin{align}\label{consistency1}
    A^2+BC&=-1 &  AB+BD&=0 \nonumber\\
    CA+DC&=0 &  CB+D^2&=-1
\end{align}
More specifically, the projection from the phase space to $\mathbb{M}$ to $\mathbb{W}$ and $\overline{\mathbb{W}}$ is described as:
\begin{align}
    \vb{K}\mqty(g\\h)&=\frac{1}{2}(1-iJ)\mqty(g\\h)=-\frac{i}{2}\mqty(
        i-A & -B \\
        -C & i-D
    )\mqty(g\\h) \nonumber\\
    &=
    -\frac{i}{2}
    \mqty(ig-Ag-Bh\\
    -Cg+ih-Dh
    )=\mqty(g_+\\h_+)
\end{align}
    
\begin{align}
    \overline{\vb{K}}\mqty(g\\h)&=\frac{1}{2}(1+iJ)\mqty(g\\h)=-\frac{i}{2}\mqty(
        i+A & B \\
        C & i+D
    )\mqty(g\\h)\nonumber\\
    &=
    -\frac{i}{2}
    \mqty(ig+Ag+Bh\\
    Cg+ih+Dh
    )=\mqty(g_-\\h_-)
\end{align}
where $\mqty(g\\h) \in \mathbb{M}$. The inner product is:
\begin{align}\label{innerProduct}
     \braket{\Psi_{\text{1-sys}}}{\Psi_{\text{1-sys}'}}     &=-i\Omega\left(\overline{\vb{K}}\lambda,\vb{K}\nu\right) \nonumber\\
     &= -\frac{1}{2}\Omega(\lambda,J\nu)-i\frac{1}{2}\Omega(\lambda,\nu)
\end{align}
The term $-\Omega(\lambda,J\nu)$ is usually denoted as $\mu(\lambda,\nu)$. Note that it is symmetric, so the following relations must be satisfied by the components of the complex structure:
\begin{align}
    \mu(\lambda,\nu)&=-\Omega(\lambda,J\nu)\nonumber\\
    &=\int_\Sigma d^3x\:(hAq+hBp-gCq-gDp)\nonumber\\
    &=-\Omega(\nu,J\lambda)\nonumber\\
    &=\int_\Sigma d^3x\:(pAg+pBh-qCg-qDh)
\end{align}
Therefore, the terms must satisfy the relations:
\begin{align}\label{consistency2}
    \int_\Sigma d^3x hAq = -\int_\Sigma d^3xqDh \nonumber\\
    \int_\Sigma d^3x hBp = \int_\Sigma d^3xpBh \nonumber\\
    \int_\Sigma d^3x gCq = \int_\Sigma d^3xqCg
\end{align}
The equations \eqref{consistency1} and \eqref{consistency2} are conditions of consistency of the components of $J$. Additionally, $\mu(\lambda,\nu)$ is the real part of the product, and here the free choice of modes was encapsulated .

\section{Relating the Fock and the Schrodinger representation.}
\subsection{The GNS construction}
The operators $\hat{\phi}$ and $\hat{\pi}$ generates elementary linear observables; therefore, it is valid to ask about the general Lie group these operators might generate. An element of this Lie group could look like:
\begin{equation}
    W[g,h]=e^{-i\int\{h(x)\hat{\phi}(x)-g(x)\hat{\pi}(x)\}dx}
\end{equation}
Regarding the exponent is $\hat{\Omega}([g,h],\cdot)$, and in order to simplify our notation $\lambda:=[g,h]$, hence:
\begin{equation}\label{rep1}
    W[g,h]=W(\lambda)=e^{-i\hat{\Omega}(\lambda,\cdot)}
\end{equation}
It is worthwhile asking about the product between two different elements of our Lie group:
\begin{multline}\label{prop1}
    W(\lambda_1)W(\lambda_2)=e^{-i\hat{\Omega}(\lambda_1,\cdot)}e^{-i\hat{\Omega}(\lambda_2,\cdot)}\\
    =e^{\frac{i}{2}\Omega(\lambda_1,\lambda_2)}e^{-i\hat{\Omega}(\lambda_1+\lambda_2,\cdot)}=e^{\frac{i}{2}\Omega(\lambda_1,\lambda_2)}W(\lambda_1+\lambda_2)
\end{multline}
where we used the Baker-Campbell-Hausdorff formula. Additionally:
\begin{equation}\label{prop2}
    W(\lambda)^*=e^{i\hat{\Omega}(\lambda,\cdot)}=e^{-i\hat{\Omega}(-\lambda,\cdot)}=W(-\lambda)
\end{equation}
And furthermore if $\lambda=0$, then:
\begin{equation}\label{prop3}
    W(0)=1
\end{equation}
The relations \eqref{prop1}, \eqref{prop2}, and \eqref{prop3} are called the Weyl relations. All operators that satisfy the Weyl relation belong to a representation of the so called \textit{Weyl algebras} $\mathcal{A}$. Quantization means associating a representation of the Weyl relations on a Hilbert space. It is easy to see that $W(\lambda)$ satisfies all the requirements of a $C^*$-algebra. Additionally,  $W(\lambda)\in R(\mathcal{A})$ are, by their own, representations of $\mathcal{A}$.

In the research conducted by \cite{Corichi2002OnTR}, it is used the GNS construction which can be stated by the following theorem:

\begin{theorem}
Let $\mathcal{A}$ be a unital $C^*$-algebra and let $\omega:\mathcal{A}\to\mathbb{C}$ be a state. Then there is a Hilbert space $\mathcal{H}$, a representation $R:\mathcal{A}\to L(\mathcal{H})$ and a vector $\ket{\Psi_0}\in \mathcal{H}$ such that,
\begin{equation}
    \omega(A)=\braket{\Psi_0}{R(A)\Psi_0}_{\mathcal{H}}
\end{equation}
Furthermore, the vector $\ket{\Psi_0}$ is cyclic. The triplet $(\mathcal{H},R,\Psi_0)$ with these properties is unique (up to unitary equivalence).
\end{theorem}

The expected value of the Weyl operator can be obtained if we apply the functional $\omega_{\text{Fock}}(\cdot)$ over $R_{\text{Fock}}(\hat{W}(g))$, which is a Fock representation of the elements of the Weyl algebra. Thus we get:
\begin{multline}\label{fockrepresentation}
    \omega(R_{\text{Fock}}(\hat{W}(g)))_{\text{Fock}}={ _{\text{Fock}}}\bra{0}R_{\text{Fock}}(W(\lambda))\ket{0}_{\text{Fock}}\\
    =\exp\left(-\frac{1}{4}\mu(\lambda,\lambda)\right)
\end{multline}
$\ket{0}_{\text{Fock}}$ is the Fock vacuum chosen by the correspondingly creation/annihilation operators. 
It is useful to note that a Schrödinger representation of the Weyl operator is as follows:
\begin{equation}
    R_{\text{Sch}}(\hat{W}(\lambda)) = e^{i\hat{\phi}[h]-i\hat{\pi}[g]}=e^{i\hat{\phi}[h]}e^{-i\hat{\pi}[g]}e^{-\frac{1}{2}[\hat{\phi}[h],\hat{\pi}[g]]}
\end{equation}
We have two operators whose action over a Hilbert space is not determined. Solving this situation requires that one of them should be fixed. Let us see which one. One of the reasons why we want to calculate the expectation value of $R_{\text{Sch}}(\hat{W}(\lambda))$ is to define normalizations of the kind ``$\int\mathcal{D}\phi\Psi^*[\phi]\Psi[\phi]=1$'' correctly. Of course, we might have chosen to integrate over $\pi$. The interpretation of $\Psi[\phi]$ is that $|\Psi[\phi]|^2$ is proportional to the probability density for the quantum field $\hat{\phi}(x,\Sigma_{t})$ to assume the value $\phi(x,\Sigma_{t})$ at the fixed surface $\Sigma_{t}$, which is parameterized by $t$. Such interpretation, automatically fixes the functional $\Psi[\phi]$ as the eigenvector of $\hat{\phi}(x,\Sigma_{t})$, namely $\hat{\phi}\Psi[\phi]=\phi\Psi[\phi]$, where I omit the hyper-surface variables. Alternatively, it can be affirmed that $\hat{\phi}$ is diagonal in the Schrödinger representation. The canonical commutator is $[\hat{\phi}[h],\hat{\pi}[g]]=i\int_{\Sigma} d^3x hg$, where we omit $t$ in $\Sigma$ henceforth. If the vacuum expectation of $R_{\text{Sch}}(\hat{W}(\lambda))$ is calculated and using the fact that expectation values should be independent of the representation, it is valid to equate the equation \eqref{fockrepresentation} with $\omega(R_{\text{Sch}}(\hat{W}(\lambda)))_{\text{Sch}}$, thus:
\begin{multline}\label{relationimportant}
    \omega(R_{\text{Sch}}(\hat{W}(\lambda)))_{\text{Sch}}\\
    =\braket{\Psi_0}{e^{i\hat{\phi}[h]}e^{-i\hat{\pi}[g]}e^{-\frac{i}{2}\int_{\Sigma} d^3x hg}\Psi_0}\\
    =e^{-\frac{i}{2}\int_{\Sigma}d^3x hg}\braket{\Psi_0}{e^{i\int_{\Sigma}d^3x\phi h}e^{-i\hat{\pi}[g]}\Psi_0}\\
    =\exp\left(-\frac{1}{4}\mu(\lambda,\lambda)\right)
\end{multline}

Through this relation, we obtain an equation that can be used to find a representation for $\hat{\pi}[g]$, which furthermore, depends of the complex structure $J$ via $\mu$. Explicitly:
\begin{multline}\label{innerFunctional}
    \braket{\Psi_0}{e^{i\int_{\Sigma}d^3x\phi h}e^{-i\hat{\pi}[g]}\Psi_0}\\
    =\exp\left(\frac{i}{2}\int_{\Sigma}d^3x hg+\frac{1}{4}\Omega(\lambda,J\lambda)\right)
\end{multline}
There is already a sketch for $\hat{\pi}[g]$, motivated by the commutation relation $[\hat{\phi}[h],\hat{\pi}[g]]=i\int_{\Sigma} d^3x hg$. In order to satisfy this relation, $\hat{\pi}[g]$ must depend on a functional derivative over $\phi$, and at most, polynomial terms on $\phi$ should be allowed. To keep the calculations from getting too complex, they work just with linear terms. The following proposition of the momentum density operator is made:
\begin{equation}\label{Momentum}
    \hat{\pi}[g] = \int_\Sigma d^3x\left(-ig\frac{\delta}{\delta\phi}+\phi (M+N)g\right)
\end{equation}
$M$ and $N$ are operators that act over elements in the solution space $\mathbb{S}$ and its cotangent space. In \cite{Corichi2002OnTR} and \cite{onTheSc} a similar momentum density operator is made but differs from \eqref{Momentum} in the term $N$. This choice is made because it is required to keep the expression as general as possible and as simple at the same time. Keeping in mind that the momentum density is in the exponent, it is important to use the BCH again\footnote{For a formal treatment of sums in the exponent, the  Zassenhaus formula works better, but in order that the commutator commutes with the other elements of the algebra, both formulas work in the same manner. Another useful relation for sums is the Suzuki–Trotter decomposition \cite{Suzuki}. }:
\begin{align}\label{BCH2}
    e^{-i\hat{\pi}[g]}&=e^{-i\int_\Sigma d^3x\phi(M+N)g-\int_\Sigma d^3xg\frac{\delta}{\delta\phi}} \nonumber\\
    &=e^{-i\int_\Sigma d^3x\phi Mg}\:e^{-\int_\Sigma d^3x(g\frac{\delta}{\delta\phi}+i\phi N g)}\nonumber\\
    &\:\:\:\times e^{-\frac{i}{2}\int_\Sigma d^3x\int_\Sigma d^3ygMg[\phi,\frac{\delta}{\delta\phi}]}\nonumber\\
    &=e^{\frac{i}{2}\int_\Sigma d^3xgMg}\:e^{-i\int_\Sigma d^3x\phi Mg}\nonumber\\
    &\:\:\:\times e^{-\int_\Sigma d^3x(g\frac{\delta}{\delta\phi}+i\phi Ng)}
\end{align}
Substituting \eqref{BCH2} in \eqref{innerFunctional}:
\begin{multline}\label{nonimpooo}
    e^{\frac{i}{2}\int_\Sigma d^3xgMg}\bra{\Psi_0} e^{i\int_{\Sigma}d^3x\phi h}\:e^{-i\int_\Sigma d^3x\phi Mg}\\
    \times e^{-\int_\Sigma d^3x(g\frac{\delta}{\delta\phi}+i\phi Ng)}\ket{\Psi_0}\\
    =\exp\left(\frac{i}{2}\int_{\Sigma}d^3x hg+\frac{1}{4}\Omega(\lambda,J\lambda)\right)
\end{multline}
Now, how can $e^{-\int_\Sigma d^3x(g\frac{\delta}{\delta\phi}+i\phi Ng)}\ket{\Psi_0}$ be calculated? 
The functional $\Psi_0$ is a sort of "free choice'', and whenever it does not come into conflict with the state induced by the annihilation operator, all the construction might be regarded as consistent. The exponential term can be written as:
\begin{multline}\label{condddd1}
    e^{-\int_\Sigma d^3x(g\frac{\delta}{\delta\phi}+i\phi Ng)}=\\\lim_{n\to\infty}\left(1-\frac{1}{n}\int_\Sigma d^3x\left[g\frac{\delta}{\delta\phi}+i\phi Ng\right]\right)^n
\end{multline}
Lets choose a functional that can satisfy the following condition:
\begin{equation}\label{eqpsi_0}
    \int_\Sigma d^3x\left[g\frac{\delta}{\delta\phi}+i\phi Ng\right]\Psi_0[\phi]=0
\end{equation}
This implies $e^{-\int_\Sigma d^3x(g\frac{\delta}{\delta\phi}+i\phi Ng)}\ket{\Psi_0}=\ket{\Psi_0}$. The states are functionals, of the form
\begin{equation}\label{sha}
    \Psi_0[\phi]=Ce^{-\frac{i}{2}\int_\Sigma d^3x\:\phi N\phi}
\end{equation}
$C$ is a normalization constant. Assume that the operator $N$ can be split in $N=N'+iN''$. An appropriate measure is needed to guaranty that the integral is well defined. Let us call it $\hat{\mu}$. Moreover, the integral shall run over the configuration space $\mathbb{S}$. Using the conditions \eqref{eqpsi_0} and \eqref{sha} in \eqref{nonimpooo} we can obtain:
\begin{multline}\label{tosimply}
    e^{\frac{i}{2}\int_\Sigma d^3xgMg}\int_\mathbb{S} d\hat{\mu} \:e^{\int_\Sigma d^3x\:\phi N''\phi}\: e^{i\int_{\Sigma}d^3x\:\phi h}\times\\ 
     e^{-i\int_\Sigma d^3x\:\phi Mg}=
    \exp\left(\frac{i}{2}\int_{\Sigma}d^3x \:hg+\frac{1}{4}\Omega(\lambda,J\lambda)\right)
\end{multline}
It is worthwhile to simplify the equation. If we take a look at \eqref{relationimportant}, one may note that this is true for any vector on the phase space, so lets us choose $\lambda=[0,h]$. If this is so, $\mu(\lambda,\lambda)=-\Omega([0,h],J[0,h])=\Omega([0,h],[Bh,Dh])=\int_\Sigma d^3xhBh$, substituting $\lambda$ and the previous result into \eqref{tosimply} leads to:
\begin{equation}\label{nonimpoe}
    \int_{\mathbb{S}} d\hat{\mu}\:e^{\int_{\Sigma_t} d^3x\:\phi N''\phi}e^{i\int_{\Sigma_t} d^3x\:\phi h}=e^{-\frac{1}{4}\int_{\Sigma_t} d^3x\:hBh}
\end{equation}
Is worthwhile to reabsorb the quadratic term in the exponential into the measure:
\begin{equation}\label{nonimpoeo}
    \int_{\mathbb{S}} d\Tilde{\mu} e^{i\int_\Sigma d^3x\:\phi h}=e^{-\frac{1}{4}\int_\Sigma d^3x\:hBh}
\end{equation}
Where $d\Tilde{\mu}=d\hat{\mu}e^{\int_{\Sigma_t}d^3x\:\phi N''\phi}$. 
The relation \eqref{nonimpoeo} is the Fourier transform of the measure $\Tilde{\mu}$. There is a theorem that links the Fourier transform with being a Gaussian measure \cite{bogachev2015gaussian}:
\begin{theorem}
A measure $\Tilde{\mu}$ on a locally convex space X is Gaussian and centered, if and only if its Fourier transform has the form:
\begin{equation}
    \chi(\Tilde{\mu})=e^{-\frac{1}{4}(h,Oh)}
\end{equation}
$O$ is a symmetric bilinear function on $X^*$ and the bilinear form and $(h,Oh)$ is positive defined.
\end{theorem}
Thus, it is concluded that $\Tilde{\mu}$ is Gaussian. The measure looks like:
\begin{equation}\label{gaussianmeasureex}
    d\Tilde{\mu}_{B^{-1}}=\mathcal{D}\phi e^{-\int_\Sigma d^3x\:\phi B^{-1}\phi}
\end{equation}
The subindex $B^{-1}$ in $d\Tilde{\mu}_{B^{-1}}$ indicates the dependence of the measure in the operator $B^{-1}$. Such result implies that from \eqref{gaussianmeasureex} and the definition of $\Tilde{\mu}$:
\begin{equation}
    d\hat{\mu}=\mathcal{D}\phi e^{-\int_{\Sigma_t} d^3x\:\phi (B^{-1}+N'')\phi}
\end{equation}
So knowing $N''$ allows us to know the integration measure. Returning to the equation \eqref{tosimply}, using the fact that $\Psi$ is a complex constant in \eqref{nonimpooo} and that we are dealing with a Gaussian measure:
\begin{multline}
    e^{\frac{i}{2}\int_\Sigma d^3xgMg}\int_{\mathbb{S}}d\Tilde{\mu}_{B^{-1}}\:e^{i\int_\Sigma d^3x\phi(h-Mg) }\\
    =\exp\left(\frac{i}{2}\int_{\Sigma}d^3x hg+\frac{1}{4}\Omega(\lambda,J\lambda)\right)
\end{multline}

Thus, the functional integral in the left hand side is:
\begin{multline}
    e^{\frac{i}{2}\int_\Sigma d^3xgMg}\:e^{-\frac{1}{4}\int_{\Sigma}d^3x(h-Mg)B(h-Mg)}\\
    =\exp\left(\frac{i}{2}\int_{\Sigma}d^3x hg+\frac{1}{4}\Omega(\lambda,J\lambda)\right)
\end{multline}
Lets dive into the term of the right hand side. The term $\Omega(\lambda,J\lambda)$ can be written using the matrix form of $J$, $\Omega([g,h],J[g,h])=-\Omega([g,h],[Ag+Bh,Cg+Dh])$:
\begin{multline}
    \Omega([g,h],[Ag+Bh,Cg+Dh])\\
    =\int_\Sigma d^3x(hAg+hBh-gCg-gDh)
\end{multline}

Hence, equating the exponents and the integrands:
\begin{multline}\label{ot}
    \frac{i}{2}gMg-\frac{1}{4}(hBh-2h(BMg)+(Mg)(BMg))\\
    =\frac{i}{2}hg-\frac{1}{4}(hAg+hBh-gCg-gDh)
\end{multline}
where the symmetry of $B$ obtained in \eqref{consistency2} was used. From here, we extract the relations:
\begin{align}\label{consistency3}
    gCg&=i2gMg-(Mg)(BMg) \\
    h(BMg)&=ihg-hAg
\end{align}
where the relations \eqref{consistency2} were used, especially the fact that $hAg=-gDh$ under integral sign. Now, from the second relation in \eqref{consistency3} it is extracted the equation:
\begin{equation}
    \boxed{M=B^{-1}(i\mathbb{I}-A)=iB^{-1}-B^{-1}A}
\end{equation}
Such operator was obtained in \cite{Corichi2002OnTR, onTheSc}. Now we will look for the operator $N''$
\subsection{Using the Schrödinger representation to find the operator N''}\label{sec:3}

The representation of the momentum density operator in the Schrödinger representation with a Gaussian measure is:
\begin{equation}\label{momentumrepres}
    \hat{\pi}[g] = \int_\Sigma d^3x\left(-ig\frac{\delta}{\delta\phi}+\phi (iB^{-1}-B^{-1}A+N)g\right)
\end{equation}
The further term found in \eqref{momentumrepres} is product of have considered a Gaussian measure. Lets see explicitly how the creation/annihilation operators look like. The creation operator in terms of $\phi$ and $\pi$ can be written as:
\begin{align}
    \hat{b}^\dagger[\textbf{K}\lambda]&=-i\hat{\Omega}(\textbf{K}\lambda,\cdot)=-i\hat{\Omega}\left(\frac{1}{2}(1-iJ)\lambda,\cdot\right)\nonumber\\
    &=\frac{1}{2}(\hat{\phi}[Cg+Dh-ih]-\hat{\pi}[Ag+Bh-ig])
\end{align}
Analogously, the annihilation operator is:
\begin{align}
    \hat{b}[\overline{\textbf{K}\lambda}]&=i\hat{\Omega}(\overline{\textbf{K}\lambda},\cdot)=i\hat{\Omega}\left(\frac{1}{2}(1+iJ)\lambda,\cdot\right)\nonumber\\
    &=\frac{1}{2}(\hat{\phi}[Cg+Dh+ih]-\hat{\pi}[Ag+Bh+ig])
\end{align}
It is straightforward to show that the annihilation operator derived from the representation of $\pi$ in the Gaussian measure is explicitly:
\begin{multline}\label{generalannihilation}
    \hat{b}_{\text{Gauss}}[\overline{\textbf{K}\lambda}] =\frac{1}{2}\int_\Sigma d^3x\: \left[(Ag+Bh+ig)i\fdv{\phi}\right.\\
    -\left.\phi N(Ag+Bh+ig)\right]
\end{multline}
Where the subindex "Gauss'' indicates the corresponding measure. 
It is simple to see that the creation operator can be written as:
\begin{multline}\label{creation}
    \hat{b}_{\text{Gauss}}^\dagger[\textbf{K}\lambda]=-\frac{1}{2}\int_\Sigma d^3x\left(2\hat{\phi}(B^{-1}(iA+1)g+ih)\right.\\
    \left.+\phi N(-i(iA+1)g+Bh)-i(Ag+Bh-ig)\frac{\delta}{\delta\phi}\right)
\end{multline}
We are free to choose the operators of creation and annihilation, in order to simplify the calculations, lets choose $\hat{b}_{\text{Gauss}}^\dagger[\textbf{K}\lambda]$ as proportional to the functional derivative.  Therefore:
\begin{equation}\label{Noperator}
    N=-2iB^{-1}
\end{equation}
Lets remember that $N=N'+iN''$. $N'$ is just a phase in $\Psi_0[\phi]$ and it can be fixed as zero. Thus:
\begin{equation}
    N''=-2B^{-1}
\end{equation}
The measure ends up being:
\begin{equation}
    d\hat{\mu}=\mathcal{D}\phi e^{\int_{\Sigma_t} d^3x\:\phi B^{-1}\phi}
\end{equation}
Which clearly is not Gaussian. Although the vacuum is Gaussian:
\begin{equation}\label{vacuum}
    \boxed{\Psi_0[\phi]=C[B]e^{-\int_\Sigma d^3x\:\phi B^{-1}\phi}}
\end{equation}
where $C[B]$ is a normalization constant. Here we make explicit the functional dependence of $B$ because the vacuum depends on its choice (the choice of the complex structure $J$).
The vacuum compensates this 'ill' definition of the measure.
Therefore, by substituting the operator \eqref{Noperator} in \eqref{momentumrepres}, we obtain:
\begin{equation}\label{mymomentumrepres}
   \hat{\pi}[g] = -i\int_\Sigma d^3x\left(g\frac{\delta}{\delta\phi}+\phi (B^{-1}-iB^{-1}A)g\right)
\end{equation}

The creation and annihilation operators are correspondingly:
\begin{align}\label{mygeneralcreation}
    \hat{b}_{\text{Gauss}}^\dagger[\textbf{K}\lambda] =\frac{i}{2}\int_\Sigma d^3x\:(Ag+Bh-ig)\fdv{}{\phi}
\end{align}
\begin{multline}\label{mygeneralannihilation}
    \hat{b}_{\text{Gauss}}[\overline{\textbf{K}\lambda}] =\frac{i}{2}\int_\Sigma d^3x\: (Ag+Bh+ig)\\
    \times \left[\fdv{\phi}+2 B^{-1} \phi\right]
\end{multline}
Alternatively, the positive and negative frequency modes can be explicitly noted:
\begin{align}\label{mygeneralcreation2}
    \hat{b}_{\text{Gauss}}^\dagger[\textbf{K}\lambda] &=\int_\Sigma d^3x\:g^+\fdv{}{\phi} \\ 
    \hat{b}_{\text{Gauss}}[\overline{\textbf{K}\lambda}] &=-\int_\Sigma d^3x\: g^-\left[\fdv{\phi}+2 B^{-1} \phi\right]
\end{align}

\section{Discussion}
A shape of the vacuum which depends on the complex structure seems more natural and compatible with the Fock description of states. For instance  the restriction of the Minkowski vacuum to the Right (or left) wedge which is highly mixed. This is consequence of the Reeh-Schliender theorem. It is expected that the restriction of the Minkowski vacuum in such a Wedge, should be written in terms of a linear combinations of Rindler's pure states. The whole Minkowski vacuum restricted to both wedges is pure and therefore, should be orthogonal to total Rindler's vacuum $\omega_R^L\otimes \omega_R^R$ \cite{arerindler}. An appropriate Schrödinger representation should manifest this fact.

I suspect that the shape of the vacuum \eqref{vacuum} respects these relations among different vacuums, even in the non-unitary case like the relation between Minkowski and Rindler displayed in (2.76) of \cite{RevModPhys.80.787}. Such constrain should be taking into account in order to reproduce theoretical results obtained in the Fock representation, like the Unruh effect or Hawking effect. Although a further analysis should be made in order to compute even the expectation values trough this representation.

\nocite{*}

\bibliography{apssamp}

\end{document}